# Experimental delayed-choice entanglement swapping

Xiao-song Ma[1,2], Stefan Zotter[1], Johannes Kofler[1,a],
Rupert Ursin[1], Thomas Jennewein[1,b], Časlav Brukner[1,3], and Anton Zeilinger[1,2,3]

[1] *Institute for Quantum Optics and Quantum Information (IQOQI), Austrian Academy of Sciences, Boltzmanngasse 3, A-1090 Vienna, Austria*
[2] *Vienna Center for Quantum Science and Technology (VCQ), Faculty of Physics, University of Vienna, Boltzmanngasse 5, A-1090 Vienna, Austria*
[3] *Faculty of Physics, University of Vienna, Boltzmanngasse 5, A-1090 Vienna, Austria*
[a] *Present Address: Max Planck Institute of Quantum Optics, Hans-Kopfermann-Str. 1, 85748 Garching/Munich, Germany*
[b] *Present Address: Institute for Quantum Computing and Department of Physics and Astronomy, University of Waterloo, 200 University Ave W., Waterloo, ON, Canada N2L3G1*

**Motivated by the question, which kind of physical interactions and processes are needed for the production of quantum entanglement, Peres has put forward the radical idea of delayed-choice entanglement swapping. There, entanglement can be "produced a posteriori, after the entangled particles have been measured and may no longer exist." In this work we report the first realization of Peres' gedanken experiment. Using four photons, we can actively delay the choice of measurement – implemented via a high-speed tunable bipartite state analyzer and a quantum random number generator – on two of the photons into the time-like future of the registration of the other two photons. This effectively projects the two already registered photons onto one definite of two mutually exclusive quantum states in which either the photons are entangled (quantum correlations) or separable (classical correlations). This can also be viewed as "quantum steering into the past".**

In the entanglement swapping[1-3] procedure, two pairs of entangled photons are produced, and one photon from each pair is sent to Victor. The two other photons from each pair are sent to Alice and Bob, respectively. If Victor projects his two photons onto an entangled state, Alice's and Bob's photons are entangled although they have never interacted or shared any common past. What might be considered as even more puzzling is Peres' idea of "delayed-choice for entanglement swapping"[4,5]. In this gedanken experiment, Victor is free to choose either to project his two photons onto an entangled state and thus project Alice's and Bob's photons onto an entangled state, or to measure them individually and then project Alice's and Bob's photons onto a separable state. If Alice and Bob measure their photons' polarization states before Victor makes his choice and projects his two photons either onto an entangled state or onto a separable state, it implies that whether their two photons are entangled (showing quantum correlations) or separable (showing classical correlations) can be defined after they have been measured.

In order to experimentally realize Peres' gedanken experiment, we place Victor's choice and measurement in the time-like future of Alice's and Bob's measurements, providing a "delayed-choice" configuration in any and all reference frames. This is accomplished by (1) proper optical delays for Victor's photons and (2) a high-speed tunable bipartite state analyzer, which (3) is controlled in real time by a quantum random number generator (QRNG)[6]. Both delay and randomness are needed to avoid the possibility that the photon pairs can "know" in advance which setting will be implemented after they are registered and can behave accordingly by producing results of a definite entangled or a definite separable state. Whether Alice's and Bob's photons can be assigned an entangled state or a separable state depends on Victor's later choice. In Peres' words: "if we attempt to attribute an objective meaning to the quantum state of a single system, curious paradoxes appear: quantum effects mimic not only instantaneous action-at-a-distance but also, as seen here, influence of future actions on past events, even after these events have been irrevocably recorded."[4]

Historically, delayed-choice entanglement swapping[4] can be seen as the fascinating consequence of quantum entanglement[1,2] emerging from combining the gedanken experiments by Bohr[7], illustrated by a double-slit setup, and Wheeler[8,9], illustrated by a Mach-Zehnder interferometer. In Bohr's gedanken experiment, he illustrated the complementarity principle, one of the most basic principles of quantum mechanics, with a double-slit apparatus. If both slits are open, the input quantum system exhibits "wave-like" behavior and shows interference on the detector screen. If only one slit is open, the system can only propagate through this slit. In this case, no interference will be observed and the system exhibits "particle-like" behavior with a well-defined path. In accordance with the complementarity principle, full interference and full path information will never be obtained simultaneously. As an explanation it is often said that any attempt to determine which path a particle takes inside an interferometer disturbs the particle and thus prevents the interference pattern from forming. From a modern point of view, however, interference patterns can arise if and only if no information about the path taken exists either on the particle itself or in the environment, regardless of whether or not an observer accesses this information.

If the choice between complementary experimental settings—one demonstrating interference, one revealing which-path information—is made in the past, an explanation of Bohr's complementarity can be given in the following way: Before the particle enters the interferometer, it "receives" information which setting has been prepared and then behaves correspondingly. For example, the two complementary settings in a photonic Mach-Zehnder configuration can be implemented by inserting or removing the output beam splitter that recombines the two interfering paths. To avoid the possibility that the photon somehow "knows" in advance whether the output beam splitter is chosen to be inserted or not, Wheeler suggested to delay this choice after the photon has passed the input beam splitter[8,9]. Many so-called "delayed-choice" experiments have been performed[10-15], including the scheme when the choice to insert or remove the output beam splitter is made at a space-time location that is space-like separated from the entrance of the photon in the interferometer[14,15]. According to Wheeler, "we have a strange inversion of the normal order of time. We, now, by moving the mirror in or out have an unavoidable effect on what we have a right to say about the already past history of that photon."[9] Evidently, even in such a delayed-choice scenario, the choice has to be made in the past light cone of the final detection of the photon.

On the other hand, delayed-choice experiments with entangled photons pave the way for new possibilities, where the choice of measurement settings on the distant photon can be made even after the other photon has been registered. This has been shown in a delayed-choice quantum eraser experiment[13], where the which-path information of one photon was erased by a later suitable measurement on the other photon. This allowed to a posteriori decide a single-particle characteristic, namely whether the already measured photon behaved as a wave or as a particle. However, while all previous delayed-choice experiments[8-15] focused on the characteristics of individual particles, delayed-choice entanglement swapping, using a four-partite entangled state, allows to a posteriori decide a two-particle characteristic and thus has qualitatively new features. Just as there is a wave-particle duality for single particles, there is an entanglement-separability duality for two particles. Entanglement and separability correspond to two mutually exclusive types of correlations between two particles.

Since Peres' proposal, there have been pioneering delayed entanglement swapping experiments[16,17]. However, none of these demonstrations implemented an active, random and delayed choice, which is required to guarantee that the photons cannot know in advance the setting of the future measurement. Thus, these experiments in principle allowed for a spatiotemporal description in which the past choice event influences later measurement events. Our experiment demonstrates entanglement-separability duality in a delayed-choice configuration via entanglement swapping for the first time. This means that it is possible to freely and a posteriori decide which type of mutually exclusive correlations two already earlier measured particles have. They can either show quantum correlations (due to entanglement) or purely classical correlations (stemming from a separable state). It can also be viewed as quantum steering[18] of bipartite states into the past. Due to the use of entanglement and active switching, it is also closely related to previous experimental tests of local realism[19-21]. Our experiment therefore implements the two important steps necessary on the way from Wheeler's to Peres's gedanken experiment: One needs to first extend Wheeler's delayed-choice experiment to the delayed-choice quantum eraser to have the possibility that a choice (for one particle) can be after the measurement (of another particle). In a second step, one has to go from the delayed-choice quantum eraser to delayed-choice entanglement swapping to be able to a posteriori decide on a two-particle characteristic and show entanglement-separability duality.

In entanglement swapping[3], two pairs of entangled photons 1&2 and 3&4 are each produced in the antisymmetric polarization-entangled Bell singlet state such that the total four-photon state has the form

$$|\Psi\rangle_{1234} = |\Psi^-\rangle_{12} \otimes |\Psi^-\rangle_{34}, \tag{1}$$

where $|\Psi^-\rangle_{12} = (|H\rangle_1|V\rangle_2 - |V\rangle_1|H\rangle_2)/\sqrt{2}$ and likewise for $|\Psi^-\rangle_{34}$. $|H\rangle_k$ ($|V\rangle_k$) denotes the horizontal (vertical) polarization state of the photon $k = 1, 2, 3, 4$. As schematically shown in Fig. 1, if Victor subjects his photons 2 and 3 to a Bell-state measurement, they become entangled. Consequently photons 1 (Alice) and 4 (Bob) also become entangled and entanglement swapping is achieved. This can be seen by rewriting Eq. (1) in the basis of Bell states of photons 2 and 3:

$$|\Psi\rangle_{1234} = \tfrac{1}{2}(|\Psi^+\rangle_{14} \otimes |\Psi^+\rangle_{23} - |\Psi^-\rangle_{14} \otimes |\Psi^-\rangle_{23} - |\Phi^+\rangle_{14} \otimes |\Phi^+\rangle_{23} + |\Phi^-\rangle_{14} \otimes |\Phi^-\rangle_{23}), \tag{2}$$

where the symmetric Bell triplet states are $|\Psi^+\rangle_{14} = (|H\rangle_1|V\rangle_2 + |V\rangle_1|H\rangle_2)/\sqrt{2}$, $|\Phi^\pm\rangle_{14} = (|H\rangle_1|H\rangle_2 \pm |V\rangle_1|V\rangle_2)/\sqrt{2}$ (and likewise for photons 2 and 3). Note that after the entanglement swapping, photons 1&2 (and 3&4) are not entangled with each other anymore, which manifests the monogamy of entanglement[22]. The entanglement swapping protocol itself has been experimentally demonstrated with various physical systems[23-29]. It is at the heart of quantum information applications and the foundations of quantum physics, and is a crucial ingredient for quantum repeaters[30,31], third-man quantum cryptography[32], loophole-free Bell tests[33] and other fundamental tests of quantum mechanics[34,35].

In our experiment, the primary events are the polarization measurements of photons 1 and 4 by Alice and Bob. They keep their data sets for future evaluation. Each of these data sets by itself and their correlations are completely random and show no structure whatsoever. The other two photons (photons 2 and 3) are delayed until after Alice and Bob's measurements, and sent to Victor for measurement. His measurement then decides the context and determines the interpretation of Alice and Bob's data. In our setup, using two-photon interference on a beam splitter combined with photon detections[36,37], Victor may perform a Bell-state measurement which projects photons 2 and 3 either onto $|\Phi^+\rangle_{23}$ or onto $|\Phi^-\rangle_{23}$. This would swap entanglement to photons 1 and 4. Instead of a Bell-state measurement, Victor could also decide to measure the polarization of these photons individually and project photons 2 and 3 either onto $|HH\rangle_{23}$ or onto $|VV\rangle_{23}$, which would result in a well-defined polarization for photons 1 and 4, i.e. a separable state. These two measurements are mutually exclusive (complementary in the Bohrian sense) in the same way as measuring particle or wave properties in an interference experiment. The choice between the two measurements is made by using a quantum random number generator (QRNG). The QRNG is based on the intrinsically random detection events of photons behind a balanced beam splitter[6] (for details see the Supplementary Information). According to Victor's choice of measurement (i.e. entangled or separable state) and his results (i.e. $|\Phi^+\rangle_{23}$, $|\Phi^-\rangle_{23}$, or $|HH\rangle_{23}$, $|VV\rangle_{23}$), Alice and Bob can sort their already recorded data into 4 subsets. They can now verify that when Victor projected his photons onto an entangled state ($|\Phi^+\rangle_{23}$ or $|\Phi^-\rangle_{23}$), each of their joint subsets behaves as if it consisted of entangled pairs of distant photons. When Victor projected his photons on a separable state ($|HH\rangle_{23}$ or $|VV\rangle_{23}$), Alice and Bob's joint subsets behave as if they consisted of separable pairs of photons. In neither case Alice and Bob's photons have communicated or interacted in the past. This indicates that quantum mechanical predictions are completely indifferent to the temporal order of Victor's choice and measurement with respect to Alice and Bob's measurements. Whether Alice and Bob's earlier measurement outcomes indicate entanglement of photons 1 and 4 strictly depends on which measurements Victor performs at a later time on photons 2 and 3.

The scheme of the experimental setup follows the proposals in Refs. (4,38) and is shown in Fig. 2. Two polarization-entangled photon pairs of photons 1&2 and 3&4 are emitted by two β-barium borate (BBO) crystals via type-II spontaneous parametric down conversion (SPDC)[39,40] in the state shown in Eq. (1). All four photons are coupled into single mode fibers. In order to fulfill the delayed-choice condition, the lengths of the fibers are chosen suitably. Photon 1 is sent to Alice and photon 4 to Bob with a 7 m fiber (35 ns), where their polarization states are measured. Photons 2 and 3 are each delayed with a 104 m fiber (520 ns) and sent to Victor, who projects photons 2 and 3 either onto an entangled state or a separable state[41]. See the Supplementary Information for details on the experimental space-time configuration.

One crucial component of our setup is Victor's high-speed tunable bipartite state analyzer (BiSA). This device is rapidly reconfigured such that it can project the photons 2 and 3 either on a product or an entangled state. It is realized with a Mach-Zehnder Interferometer and consists of two 50:50 beam splitters, mirrors, and most importantly two eighth-wave plates (EWP) and electro-optic modulators (EOM). The combination of EWP and EOMs acts as switchable quarter-wave plate, where the QRNG determines whether it acts as a quarter wave plate oriented along $45^{\circ}$ or whether it has no effect. We define the phase of the interferometer to be zero when all the photons that entered from input b exited into b'' (Fig. 2), which is also the phase locking point of the interferometer. For detailed information on the tunable BiSA, refer to the caption of Fig. 2, the Supplementary Information and Ref. (42).

The two complementary measurements are realized in the following ways: The Bell-state measurement (BSM) corresponds to turning on the switchable quarter-wave plates. In this case, the phase of the interferometer is $\pi/2$ and the interferometer acts as a 50/50 beam splitter. Therefore, the two photons interfere and are projected onto a Bell state by polarization-resolving single-photon detections. The separable-state measurement (SSM) corresponds to turning off the switchable quarter-wave plates. The phase of the interferometer is 0 and the interferometer acts as a 0/100 beam splitter, i.e. a fully reflective mirror. Therefore, the two photons do not interfere and are projected onto a separable state by polarization-resolving single-photon detections. When used for a BSM, our BiSA can project onto two of the four Bell states, namely onto $|\Phi^+\rangle_{23} = (|HH\rangle_{23} + |VV\rangle_{23})/\sqrt{2}$ (both detectors in b'' firing or both detectors in c'' firing) and $|\Phi^-\rangle_{23} = (|HH\rangle_{23} - |VV\rangle_{23})/\sqrt{2}$ (one photon in b'' and one in c'' with the same polarization). When an SSM is made, we look at coincidences between b'' and c'' with the same polarization (as in the BSM case for $|\Phi^-\rangle_{23}$) and the resultant projected states are $|HH\rangle_{23}$ and $|VV\rangle_{23}$. Victor's detector coincidences with one horizontal and one vertical photon in spatial modes b'' and c'' indicate the states $|HV\rangle_{23}$ and $|VH\rangle_{23}$, which are always discarded because they are separable states independent of Victor's choice and measurement.

For each successful run (a 4-fold coincidence count), not only Victor's measurement event happens 485 ns later than Alice and Bob's measurement events, but Victor's choice happens in an interval of 14 ns to 313 ns later than Alice and Bob's measurement events. Therefore, independent of the reference frame, Victor's choice and measurement are in the future light cones of Alice and Bob's measurements. Given the causal structure of special relativity, i.e. that past events can influence (time-like) future events but not vice versa, we explicitly implemented the delayed-choice scenario as described by Peres. Only after Victor's measurement, we can assert the quantum states shared by Alice and Bob. Our experiment relies on the assumption of the statistical independence of the QRNG from other events, in particular Alice and Bob's measurement results. Note that in a conspiratorial fashion, Victor's choice might not be free but always such that he chooses a separable-state measurement whenever Alice and Bob's pair is in a separable-state, and he chooses a Bell-state measurement whenever their pair is in an entangled state. This would preserve the viewpoint that in every single run Alice and Bob do receive a particle pair in a definite separable or a definite entangled state. A possible improvement of our set-up would be space-like separation of Victor's choice event and the measurement events of Alice and Bob to further strength the assumption of the mutual independence of these events.

For each pair of photons 1&4, we record the chosen measurement configurations and the 4-fold coincidence detection events. All raw data are sorted into four subensembles in real time according to Victor's choice and measurement results. After all the data had been taken, we calculated the polarization correlation function of photons 1 and 4. It is derived from their coincidence counts of photons 1 and 4 conditional on projecting photons 2 and 3 to $|\Phi^-\rangle_{23} = (|HH\rangle_{23} - |VV\rangle_{23})/\sqrt{2}$ when the Bell-state measurement was performed, and to $|HH\rangle_{23}$ or $|VV\rangle_{23}$ when the separable state measurement was performed. The normalized correlation function $E(j)$ between two photons is defined as:

$$E(j) = \frac{C(j,j)+C(j^\perp,j^\perp)-C(j^\perp,j)-C(j,j^\perp)}{C(j,j)+C(j^\perp,j^\perp)+C(j^\perp,j)+C(j,j^\perp)}, \quad (3)$$

where $j/j^\perp$ stands for horizontal/vertical ($|H\rangle/|V\rangle$) or plus/minus ($|+\rangle/|-\rangle$, with $|\pm\rangle = (|H\rangle \pm |V\rangle)/\sqrt{2}$), or right/left ($|R\rangle/|L\rangle$, with $|R\rangle = (|H\rangle + i|V\rangle)/\sqrt{2}$ and $|L\rangle = (|H\rangle - i|V\rangle)/\sqrt{2}$) circular polarization. $C(j,j^\perp)$ is the coincidence counts under the setting of $(j,j^\perp)$. According to the definition of $E(j)$ in Eq. (3), two photons are correlated/uncorrelated/anti-correlated in the basis $j/j^\perp$ when $E(j) = 1/0/-1$. In Fig. 3, we show the correlation functions of photons 1 and 4 in these three mutually unbiased bases derived from the measurement results. Note that the reason why we use one specific entangled state but both separable states to compute the correlation function is that the measurement solely depends on the settings of the EOMs in the BiSA. Then the same coincidence counts (*HH* and *VV* combinations of Victor's detectors) are taken for the computation of the correlation function of photons 1 and 4. These counts can belong to Victor obtaining the entangled state $|\Phi^-\rangle_{23}$ in a BSM or the states $|HH\rangle_{23}$ and $|VV\rangle_{23}$ in an SSM. For the details, see Supplementary information.

We quantified the quality of the experimentally obtained states $\hat{\rho}_{\mathrm{exp}}$ via the fidelity defined as $F(\hat{\rho}_{\mathrm{exp}}, |\mathrm{out}\rangle_{\mathrm{id}}) = \mathrm{Tr}(\hat{\rho}_{\mathrm{exp}}|\mathrm{out}\rangle_{\mathrm{id}}\langle\mathrm{out}|)$, which is the overlap of $\hat{\rho}_{\mathrm{exp}}$ with the ideally expected output state $|\mathrm{out}\rangle_{\mathrm{id}}$. The state fidelity of the Bell state can be decomposed into averages of local measurements in terms of Pauli $\sigma$ matrices[43,44], such as

$$F(\hat{\rho}_{\mathrm{exp}}, |\Phi^-\rangle) = \mathrm{Tr}(\hat{\rho}_{\mathrm{exp}}|\Phi^-\rangle\langle\Phi^-|) = \tfrac{1}{4}\mathrm{Tr}[\hat{\rho}_{\mathrm{exp}}(\hat{\mathrm{I}} + \hat{\sigma}_z\hat{\sigma}_z + \hat{\sigma}_y\hat{\sigma}_y - \hat{\sigma}_x\hat{\sigma}_x)] \quad (4)$$

where $\hat{\mathrm{I}}$ is the identity operator for both photons. An entanglement witness is also employed to characterize whether entanglement existed between the photons. It is defined as[43,45] $\widehat{W}(|\mathrm{out}\rangle_{\mathrm{id}}) = \frac{1}{2}\hat{\mathrm{I}} - |\mathrm{out}\rangle_{\mathrm{id}}\langle\mathrm{out}|$. A negative expectation value of this entanglement witness operator, $W(\hat{\rho}_{\mathrm{exp}}, |\mathrm{out}\rangle_{\mathrm{id}}) = \mathrm{Tr}[\hat{\rho}_{\mathrm{exp}}\hat{W}(|\mathrm{out}\rangle_{\mathrm{id}})] = \frac{1}{2} - F(\hat{\rho}_{\mathrm{exp}}, |\mathrm{out}\rangle_{\mathrm{id}})$, is a sufficient condition for entanglement.

Fig. 3A shows that when Victor performs the Bell-state measurement and projects photons 2 and 3 onto $|\Phi^-\rangle_{23}$, this swaps the entanglement, which is confirmed by significant correlations of photons 1 and 4 in all three bases. The state fidelity $F(\hat{\rho}_{\mathrm{exp}}, |\Phi^-\rangle_{14})_{\mathrm{BSM}}$ is 0.681 ± 0.034 and the entanglement witness value $W(\hat{\rho}_{\mathrm{exp}}, |\Phi^-\rangle_{14})_{\mathrm{BSM}}$ is –0.181 ± 0.034, which demonstrates entanglement between photons 1 and 4 with more than 5 standard deviations. The imperfections of the results are mainly due to the higher order emissions from SPDC, as explained in Supplementary Information. Note that while Victor, via the QRNG, can choose to make a Bell-state measurement (with possible outcomes $|\Phi^-\rangle_{23}$ and $|\Phi^+\rangle_{23}$), he cannot choose the specific outcome. If photons 2&3 are projected onto the entangled state $|\Phi^+\rangle_{23}$, photons 1&4 are projected to $|\Phi^+\rangle_{14}$ according to Eq. (2). These results are summarized in the Supplementary Information.

On the other hand, when Victor performs the separable-state measurement on photons 2 and 3 and does not swap entanglement, the correlation only exists in the $|H\rangle/|V\rangle$ basis and vanishes in the $|+\rangle/|-\rangle$ and $|R\rangle/|L\rangle$ bases, as shown in Fig. 3B. This is a signature that photons 1 and 4 are not entangled but in a separable state. The state fidelity $F(\hat{\rho}_{\mathrm{exp}}, |\Phi^-\rangle_{14})_{\mathrm{SSM}}$ is 0.421 ± 0.029 and the entanglement witness value $W(\hat{\rho}_{\mathrm{exp}}, |\Phi^-\rangle_{14})_{\mathrm{SSM}}$ is 0.078 ± 0.029, which is consistent with a separable state. Further choice-dependent results of photons 2 and 3 were obtained, as summarized in Table 1.

When Victor performed the separable-state measurement on photons 2 and 3, we find that entanglement between photons 1&2 and between photons 3&4 remained. These entanglements vanished when Victor performed the Bell-state measurement on photons 2 and 3. This is consistent with the entanglement monogamy relation. We observe that the state fidelities $F(\hat{\rho}_{\mathrm{exp}}, |\Psi^-\rangle_{12})_{\mathrm{SSM}}$ and $F(\hat{\rho}_{\mathrm{exp}}, |\Psi^-\rangle_{34})_{\mathrm{SSM}}$ of photons 1&2 and photons 3&4 are 0.908 ± 0.016 and 0.864 ± 0.019, respectively, when the separable-state measurement is performed, confirming entanglement of photons 1&2 and photons 3&4. When the Bell-state measurement is performed, $F(\hat{\rho}_{\mathrm{exp}}, |\Psi^-\rangle_{12})_{\mathrm{BSM}}$ and $F(\hat{\rho}_{\mathrm{exp}}, |\Psi^-\rangle_{34})_{\mathrm{BSM}}$ are 0.301 ± 0.039 and 0.274 ± 0.039, respectively, consistent with photons 1&2 and photons 3&4 being in separable states. This is also summarized in Table 1.

With our ideal realization of the delayed-choice entanglement swapping gedanken experiment, we have demonstrated a generalization of Wheeler's "delayed-choice" tests, going from the wave-particle duality of a single particle to the entanglement-separability duality of two particles[41]. Whether these two particles are entangled or separable has been decided after they have been measured. If one views the quantum state as a real physical object, one could get the seemingly paradoxical situation that future actions appear as having an influence on past and already irrevocably recorded events. However, there is never a paradox if the quantum state is viewed as to be no more than a "catalogue of our knowledge"[2]. Then the state is a probability list for all possible measurement outcomes, the relative temporal order of the three observer's events is irrelevant and no physical interactions whatsoever between these events, especially into the past, are necessary to explain the delayed-choice entanglement swapping. What, however, is important is to relate the lists of Alice, Bob and Victor's measurement results. On the basis of Victor's measurement settings and results, Alice and Bob can group their earlier and locally totally random results into subsets which each have a different meaning and interpretation. This formation of subsets is independent of the temporal order of the measurements. According to Wheeler, Bohr said: "No elementary phenomenon is a phenomenon until it is a registered phenomenon."[7,8] We would like to extend this by saying: "Some registered phenomena do not have a meaning unless they are put in relationship with other registered phenomena."

**Acknowledgements**

We are grateful to N. Tetik and A. Qarry for help during the early stages of the experiment, and M. Aspelmeyer and P. Walther for fruitful discussions. We acknowledge support from the European Commission, Q-ESSENCE (No. 248095), ERC Senior Grant (QIT4QAD), JTF, as well as SFB-FOQUS and the Doctoral Program CoQuS of the Austrian Science Fund (FWF).

**References**

1. A. Einstein, B. Podolsky, and N. Rosen, Can quantum-mechanical description of physical reality be considered complete? *Phys. Rev.* **47**, 777-780 (1935).
2. E. Schrödinger, Die Gegenwärtige Situation in der Quantenmechanik. *Naturwissenschaften* **23**, 807-812; 823-828; 844-849 (1935), English translation in Proceedings of the American Philosophical Society, 124 (1980) reprinted in in Quantum Theory and Measurement, J. A. Wheeler, W. H. Zurek, Eds. (Princeton Univ. Press, Princeton, NJ, 1984), pp. 152–167.
3. M. Žukowski, A. Zeilinger, M. A. Horne, A. K. Ekert, ‘‘Event-ready-detectors’’ Bell experiment via entanglement swapping. *Phys. Rev. Lett.* **71**, 4287-4290 (1993).
4. A. Peres, Delayed choice for entanglement swapping. *J. Mod. Opt.* **47**, 139-143 (2000).
5. O. Cohen, Counterfactual entanglement and nonlocal correlations in separable states. *Phys. Rev. A* **60**, 80-84 (1999).
6. T. Jennewein, U. Achleitner, G. Weihs, H. Weinfurter, A. Zeilinger, A fast and compact quantum random number generator. *Rev. Sci. Instrum.* **71**, 1675-1680 *(2*000).
7. N. Bohr, in Quantum Theory and Measurement, J. A. Wheeler and W. H. Zurek, Eds. (Princeton University Press, 1984), pp.9-49.
8. J. A. Wheeler, in Mathematical Foundations of Quantum Theory (Academic, New York, 1978), pp. 9–48.
9. J. A. Wheeler, in Quantum Theory and Measurement, J. A. Wheeler, W. H. Zurek, Eds. (Princeton Univ. Press, Princeton, NJ, 1984), pp. 182–213.
10. C. O. Alley, O. G. Jacubowicz, W. C. Wickes, in Proceedings of the Second International Symposium on the Foundations of Quantum Mechanics, H. Narani, Ed. (Physics Society of Japan, Tokyo, 1987), pp. 36–47.
11. T. Hellmut, H. Walther, A. G. Zajonc, W. Schleich, Delayed-choice experiments in quantum interference. *Phys. Rev. A* **35**, 2532-2541 (1987).
12. J. Baldzuhn, E. Mohler, W. Martienssen, A wave-particle delayed-choice experiment with a single-photon state. *Z. Phys. B* **77**, 347-352 (1989).
13. Y-H. Kim, R. Yu, S. Kulik, Y. Shih, M. O. Scully, Delayed „choice“ quantum eraser. *Phys. Rev. Lett.* **84**, 1-4 (2000).
14. V. Jacques *et al.*, Experimental realization of Wheeler's delayed-choice gedanken experiment. *Science* **315**, 966-968 (2007).
15. V. Jacques *et al.*, Delayed-Choice Test of Quantum Complementarity with Interfering Single Photons. *Phys. Rev. Lett.* **100,** 220402 (2008).
16. T. Jennewein, G. Weihs, J.-W. Pan, A. Zeilinger, Experimental Nonlocality Proof of Quantum Teleportation and Entanglement Swapping. *Phys. Rev. Lett.* ***88***, 017903 (2001).
17. F. Sciarrino, E. Lombardi, G. Milani, F. De Martini, Delayed-choice entanglement swapping with vacuum–one-photon quantum states. *Phys. Rev. A* **66**, 024309 (2002).
18. E. Schrödinger, Discussion of probability relations between separated systems. *Proc. Camb. Phil. Soc.* **31**, 555-563 (1935).
19. A. Aspect, J. Dalibard, G. Roger, Experimental test of Bell's inequalities using time-varying analyzers. *Phys. Rev. Lett.* **49**, 1804-1807 (1982).
20. G. Weihs *et al.* Violation of Bell's inequality under strict Einstein locality conditions. *Phys. Rev. Lett.* **81**, 5039-5043 (1998).
21. T. Scheidl *et al*. Violation of local realism with freedom of choice. *Proc. Natl. Acad. Sci. USA* **107**, 19708-19713 (2010).
22. V. Coffman, J. Kundu, W. K. Wootters, Distributed entanglement. *Phys. Rev. A*. **61**, 052306 (1993)**.**
23. J.-W. Pan, D. Bouwmeester, H. Weinfurter, A. Zeilinger, Experimental entanglement swapping: entangling photons that never interacted. *Phys. Rev. Lett.* **80**, 3891-3894 (1998).
24. M. Riebe *et al.*, Deterministic quantum teleportation with atoms. *Nature* **429**, 734-737 (2004).
25. M. D. Barrett *et al.,* Deterministic quantum teleportation of atomic qubits. *Nature* **429**, 737 (2004).
26. D. N. Matsukevich, P. Maunz, D. L. Moehring, S. Olmschenk, C. Monroe, Bell inequality violation with two remote atomic qubits*. Phys. Rev. Lett.* **100**, 150404 (2008).
27. M. Halder *et al.*, Entangling independent photons by time measurement. *Nature Phys.* **3**, 692-695 (2007).
28. Z.-S. Yuan *et al.*, Experimental demonstration of a BDCZ quantum repeater node. *Nature* **454**, 1098-1101 (2008).
29. R. Kaltenbaek. R. Prevedel, M. Aspelmeyer, A. Zeilinger, High-fidelity entanglement swapping with fully independent

sources. *Phys.Rev. A.* **79**, 040302 (2009).

30. H.-J. Briegel, W. Duer, J. I. Cirac, P. Zoller, Quantum repeaters: The role of imperfect local operations in quantum communication. *Phys. Rev. Lett.* **81**, 5932–5935 (1998).
31. L.-M. Duan, M. D. Lukin, J. I. Cirac, P. Zoller, Long-distance quantum communication with atomic ensembles and linear optics. *Nature* **414**, 413–418 (2001).
32. Y. -A. Chen *et al.*, Experimental quantum secret sharing and third-man quantum cryptography. *Phys. Rev. Lett.* **95**, 200502 (2005).
33. C**.** Simon, W. T. M. Irvine, Robust long-distance entanglement and a loophole-free Bell test with ions and photons. *Phys. Rev. Lett.* **91**, 110405 (2003).
34. D. M. Greenberger, M. Horne, A. Zeilinger, Bell theorem without inequalities for two particles. I. Efficient detectors. *Phys. Rev. A.* **78***,* 022110 (2008)**.**
35. D. M. Greenberger, M. Horne, A. Zeilinger, M. Žukowski, Bell theorem without inequalities for two particles. II. Inefficient detectors. *Phys. Rev. A.* **78**, 022111 (2008)**.**
36. C. K. Hong, Z. Y. Ou, L. Mandel, Measurement of subpicosecond time intervals between two photons by interference. *Phys. Rev. Lett.* **59**, 2044-2046 (1987).
37. K. Mattle, H. Weinfurter, P. G. Kwiat, A. Zeilinger, Dense Coding in Experimental Quantum Communication. *Phys. Rev. Lett.* **76**, 4656-4659 (1996).
38. T. Jennewein, M. Aspelmeyer, Č. Brukner**,** A. Zeilinger, Experimental proposal of switched delayed-choice for entanglement swapping. *Int. J. of Quant. Info.* **3**, 73 (2005).
39. P. G. Kwiat *et al.*, New high-intensity source of polarization-entangled photon pairs. *Phys. Rev. Lett.* **75,** 4337-4341 (1995).
40. M. Zukowski, A. Zeilinger, H. Weinfurter, Entangling photons radiated by independent pulsed sources. *Ann. N.Y. Acad. Sci.* **755**, 91-102 (1995).
41. Č. Brukner, M. Aspelmeyer, A. Zeilinger, Complementarity and information in delayed-choice for entanglement swapping. *Found. Phys.* **35**, 1909 (2005).
42. X.-S. Ma *et al.,* A high-speed tunable beam splitter for feed-forward photonic quantum information processing. *Opt. Express*, **19**, 22723-22730 (2011).
43. O. Gühne *et al.*, Detection of entanglement with few local measurements. *Phys. Rev. A* **66**, 062305 (2002).
44. Q. Zhang *et al.*, Experimental quantum teleportation of a two-qubit composite system. *Nature Phys.* **2**, 678-682 (2006).
45. O. Gühne, G. Toth, Entanglement detection. *Phys. Rep.* **474**, 1 (2009).

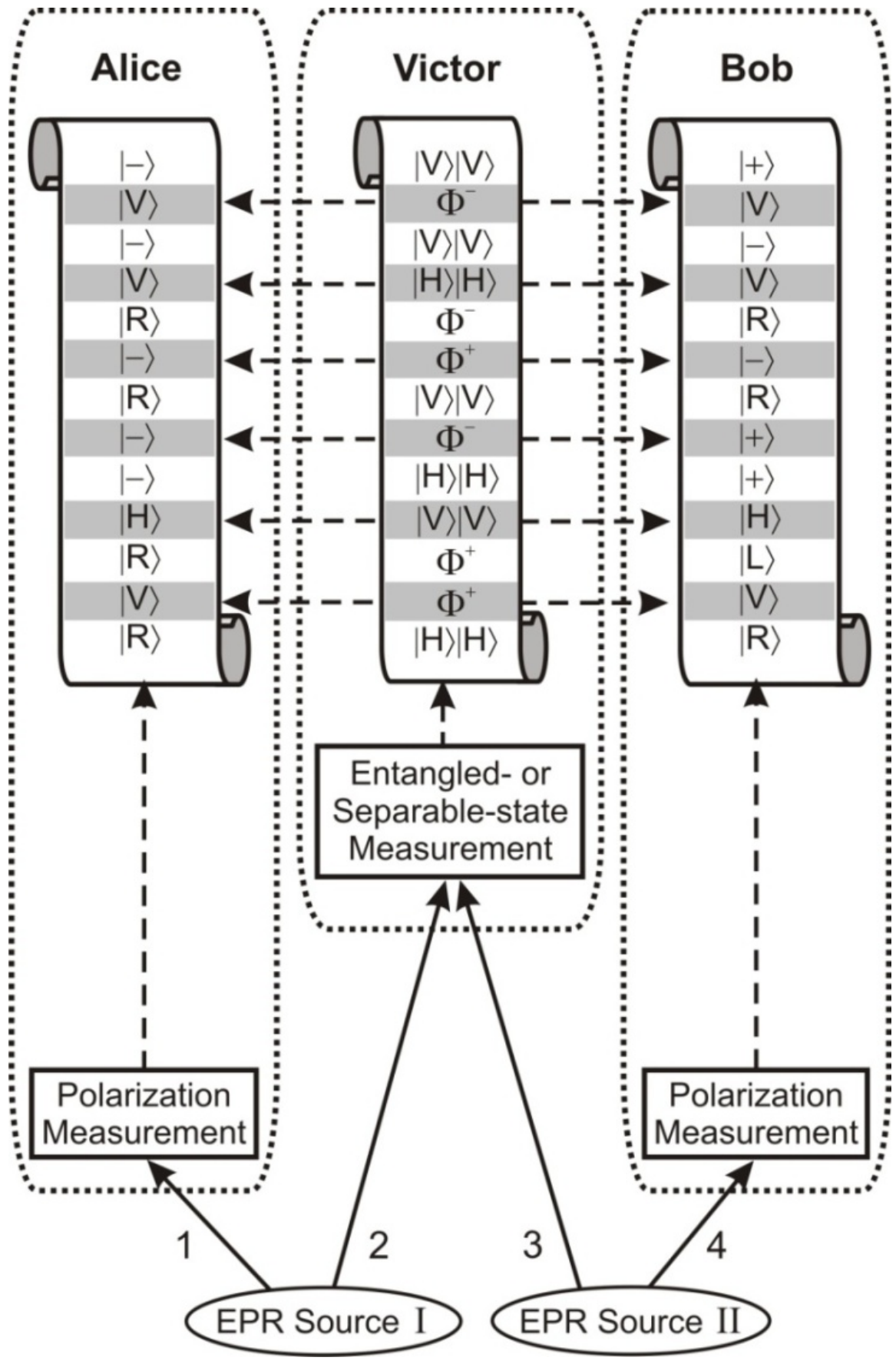

Figure 1: The concept of delayed-choice entanglement swapping. Two entangled pairs of photons 1&2 and 3&4 are produced in the state $|\Psi^-\rangle_{12} \otimes |\Psi^-\rangle_{34}$ in the EPR sources I and II, respectively. At first, Alice and Bob perform polarization measurements on photons 1 and 4, choosing freely the polarization analysis basis among three mutually unbiased bases ($|H\rangle/|V\rangle$, $|R\rangle/|L\rangle$, $|+\rangle/|-\rangle$), and record the outcomes. Photons 2 and 3 are sent to Victor who then subjects them to either an entangled or a separable-state measurement, projecting them randomly onto one of two possible Bell states ($|\Phi^+\rangle_{23}$ or $|\Phi^-\rangle_{23}$) or one of two separable states ($|HH\rangle_{23}$ or $|VV\rangle_{23}$). Victor records the outcome and keeps it to himself. This procedure projects photons 1 and 4 onto a corresponding entangled ($|\Phi^+\rangle_{14}$ or $|\Phi^-\rangle_{14}$) or separable state ($|VV\rangle_{14}$ or $|HH\rangle_{14}$), respectively. According to Victor's choice and his results, Alice and Bob can sort their already recorded data into subsets and can verify that each subset behaves as if it consisted of either entangled or separable pairs of distant photons, which have neither communicated nor interacted in the past.

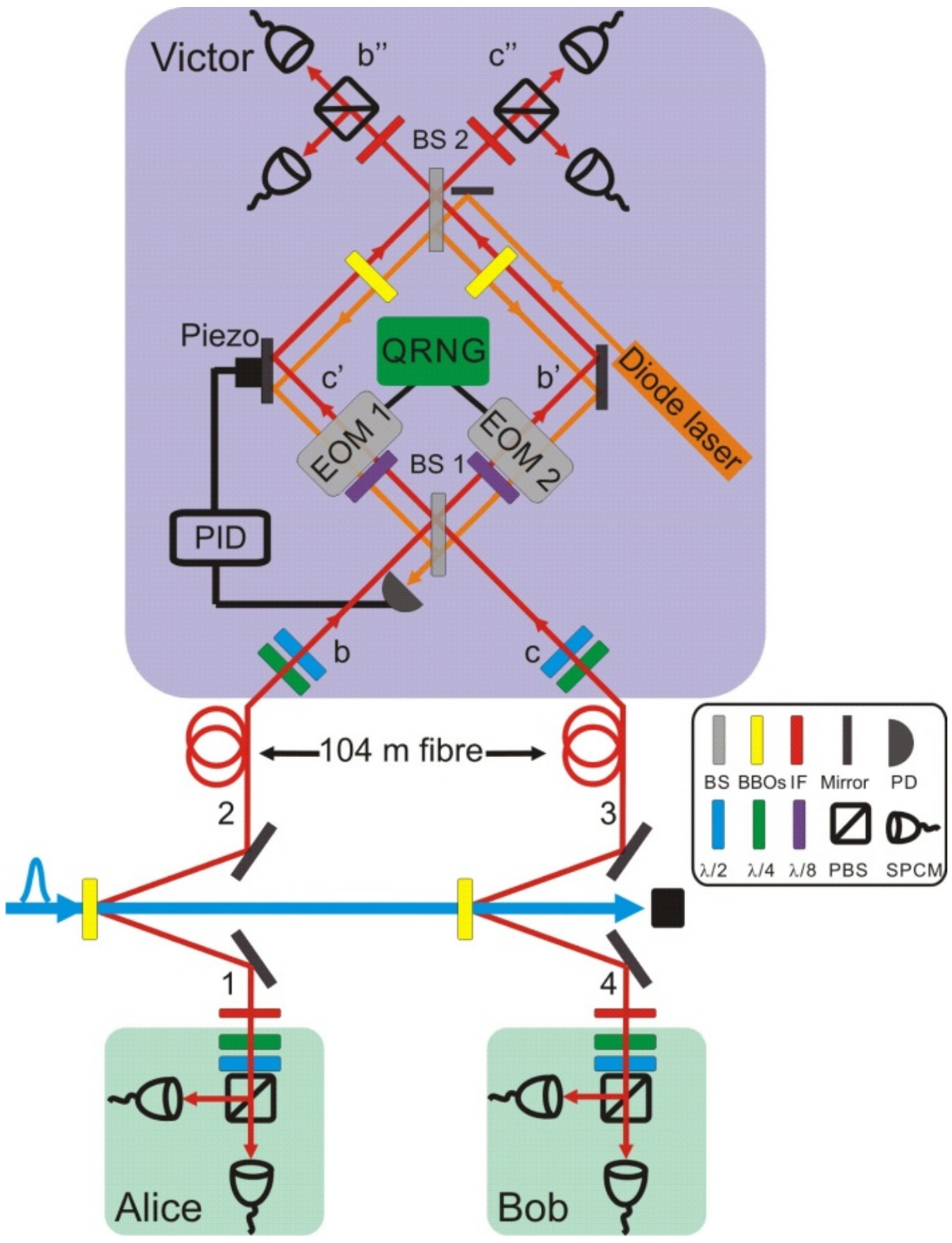

Figure 2: Experimental setup. A pulsed ultraviolet laser beam with a central wavelength of 404 nm, a pulse duration of 180 fs, and a repetition rate of 80 MHz successively passes through two β-barium borate (BBO) crystals to generate two polarization entangled photon pairs (photons 1&2 and photons 3&4) via type-II spontaneous parametric down-conversion[39,40]. Single-mode fibers and interference filters (IF) are used to clean their spatial and spectral modes. We use the interference filters with 1 nm (3 nm) bandwidth centered around 808 nm for photons 2 and 3 (photons 1 and 4). Photons 1 and 4 are directly subject to the polarization measurements performed by Alice and Bob (green blocks). Photons 2 and 3 are each delayed with 104 m single mode fiber and then coherently overlapped on the tunable bipartite state analyzer (BiSA) (purple block). The single mode fiber coupler of photon 2 is mounted on step motors and used to compensate the time delay for the interference at the tunable BiSA. An active phase stabilization system is employed in order to compensate the phase noise in the tunable BiSA, which is composed with an auxiliary power-stabilized diode laser, a photon detector (PD) and a ring piezo-transducer controlled by an analogue proportional-integral-derivative (PID) regulator. Two pairs of cross oriented BBO crystals (BBOs3 and BBOs4) are placed in each arm of the MZI in order to compensate the unwanted birefringence. On each spatial mode, we employ the combination of a half-wave plate (λ/2), a quarter-wave plate (λ/4) and a polarizing beam splitter (PBS) for measuring the pair-wise correlations between different photons in different polarization bases. The four-fold coincidence counts is about 0.016 Hz. See Supplementary Information for details.

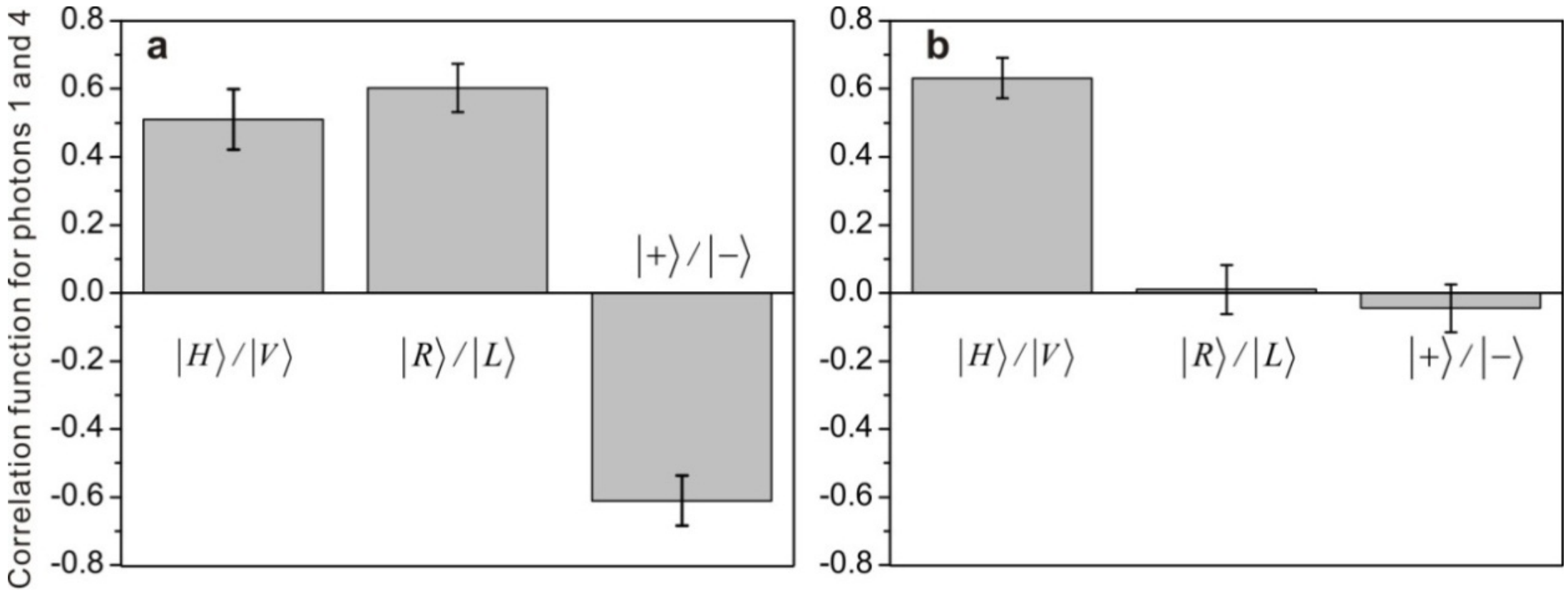

Figure 3: Experimental results: correlation function between photons 1 and 4 for the three mutually unbiased bases ($|H\rangle/|V\rangle$, $|R\rangle/|L\rangle$, $|+\rangle/|-\rangle$). Victor subjects photons 2 and 3 to either (**a**) an entangled or (**b**) a separable-state measurement. These results are obtained via coincidence counts of photons 1 and 4, conditioned on the coincidence of same polarization and different spatial output modes of photons 2 and 3 (b'' and c'' in Fig. 2). (**a**) When Victor performs a Bell-state measurement and finds photons 2 and 3 in the state $|\Phi^-\rangle_{23} = (|HH\rangle_{23} - |VV\rangle_{23})/\sqrt{2}$, entanglement is swapped to photons 1 and 4. This is confirmed by all three correlation functions being of equal magnitude (within statistical error) and their absolute sum exceeding 1. (**b**) When Victor performs a separable-state measurement and finds photons 2 and 3 in either the state $|HH\rangle_{23}$ or $|VV\rangle_{23}$, entanglement is not swapped. This is confirmed by only the correlation function in the $|H\rangle/|V\rangle$ basis being significant while the others vanish. The experimentally obtained correlation functions of photons 1 and 4 in the $|H\rangle/|V\rangle$, $|R\rangle/|L\rangle$, $|+\rangle/|-\rangle$ bases are 0.511 ± 0.089, 0.603 ± 0.071, –0.611 ± 0.074 respectively for case (**a**) and 0.632 ± 0.059, 0.01 ± 0.072, –0.045 ± 0.070 respectively for case (**b**). While entangled states can show maximal correlations in all three bases (the magnitude of all correlation functions equals 1 ideally), separable states can be maximally correlated (ideal correlation function 1) only in one basis, the others being 0. The uncertainties represent plus/minus one standard deviation deduced from propagated Poissonian statistics.

| Photon pairs | Bell-state measurement by Victor | | Separable-state measurement by Victor | |
|---|---|---|---|---|
| | State fidelity | Entanglement witness value | State fidelity | Entanglement witness value |
| Photons 2 and 3 in $\|\Phi^-\rangle$ | 0.645 ± 0.031 | –0.145 ± 0.031 | 0.379 ± 0.026 | 0.120 ± 0.026 |
| Photons 1 and 4 in $\|\Phi^-\rangle$ | 0.681 ± 0.034 | –0.181 ± 0.034 | 0.421 ± 0.029 | 0.078 ± 0.029 |
| Photons 1 and 2 in $\|\Psi^-\rangle$ | 0.301 ± 0.039 | 0.199 ± 0.039 | 0.908 ± 0.016 | –0.408 ± 0.016 |
| Photons 3 and 4 in $\|\Psi^-\rangle$ | 0.274 ± 0.039 | 0.226 ± 0.039 | 0.864 ± 0.019 | –0.364 ± 0.019 |

Table 1: Results of the state fidelities and the expectation values of the entanglement witness operator for different pairs of photons with delayed-choice condition. A negative witness value (or, equivalently, a state fidelity above ½) corresponds to an entangled state (shown in orange). When Victor performs a Bell-state measurement and photons 2&3 are found in the state $|\Phi^-\rangle_{23}$, then photons 1&4 were in the entangled state $|\Phi^-\rangle_{14}$, i.e. the entanglement was swapped. When Victor performs a separable-state measurement, projecting photons 2&3 on the mixture of $|HH\rangle_{23}$ or $|VV\rangle_{23}$, correlations between measurement results on photon pairs 1&2 and 3&4 show that these pairs were entangled in the states $|\Psi^-\rangle_{12}$ and $|\Psi^-\rangle_{34}$, i.e. the entanglement was not swapped. This can be obtained by evaluations of the pair-wise correlations between different photons. The uncertainties represent plus/minus one standard deviation deduced from propagated Poissonian statistics.

## Supplementary Information

### High-speed tunable bipartite state analyzer

We use a Mach-Zehnder interferometer (MZI) to realize the high-speed tunable bipartite state analyzer (BiSA). It consists of two 50:50 beam splitters, mirrors and most importantly two electro-optical modulators. The whole interferometer is built in a box enclosed with acoustic and thermal isolation materials in order to stabilize the phase passively. Additional to the passive stabilization, active phase stabilization is also employed as described in the text and captions of Fig. 2. Because of the short coherence time of the down-converted photon defined by the transmission bandwidth (1 nm) of the interference filter, it requires accurate path length adjustment of the two MZI arms. We minimize the path length difference by maximizing the single photon interference visibility when one input is blocked. Inside the tunable BiSA, two additional pairs of cross oriented BBO crystals (BBOs3 and BBOs4) are placed in each arm of the MZI in order to compensate the birefringence induced by the beam splitters.

As to electro optical modulators (EOM), we use Pockels Cells (PoC) consisting of two 4x4x10 $mm^3$ Rubidium Titanyl Phosphate (RTP) crystals. We align the optical axes of the RTP crystals to 45° for both EOM1 and EOM2. Additionally, we place two eighth-wave plates (EWP) with their optical axes oriented parallel (in front of the EOM1) and orthogonal (in front of the EOM2) to the axis of the RTP crystals, respectively. Applying positive eighth-wave voltage (+EV) makes the EOM1 act as an additional eighth-wave plate. Given the fact that the optical axis of EWP1 is oriented parallel to that of RTP crystals, the overall effect is the one of a quarter-wave plate (QWP) at 45°. On the other hand, applying negative eighth-wave voltage (–EV) makes the EOM1 compensate the action of the EWP1, such that there is no overall effect. Since the optical axis of EWP2 is oriented orthogonal to that of the RTP crystal in EOM2, the overall effect is the one of a QWP at –45° by applying –EV and identity by applying +EV.

With opposite voltages on EOM1 and EOM2, we realize a $\pi/2$ phase change of the MZI (corresponding to Bell-state measurement) when EOM1 is applied with +EV and EOM2 with –EV, and no phase change (corresponding to separable state measurement) when EOM1 is applied with -EV and EOM2 with +EV.

A self-built field programmable gate array based logic samples the random bit sequence from the quantum random number generator (QRNG) and delivers the required pulse sequence for the PoC driver. Corresponding to the random bit value “1” (“0”) a phase change of the MZI of $\pi/2$ (0) is applied. A certain setting is not changed until the occurrence of an opposite trigger signal. Since our QRNG is balanced within the statistical uncertainties, +QV and –QV are applied equally often approximately. Therefore, the mean field in the PoC is zero which allows continuous operation of the PoC without damaging the crystals. For the fast and optimal operation of the PoC, a sampling frequency of 2 MHz is chosen. We set the on-time of both EOMs to be 299 ns, which guarantees that the choice is delayed by more than 14 ns. The duty cycle of our tunable BiSA is about 60%. For detailed information on this apparatus refer to Ref. (1).

### Bipartite state analyzer for Bell-state measurement

When the phase of the MZI is $\pi/2$, the state evolves in the following two ways: (a) If we detect a coincidence in the same spatial mode b’’/c’’ (as shown in Fig. 2 of the main text) but with different polarization in the $|H\rangle/|V\rangle$ basis, photons 2 and 3 are projected onto $|\Phi^+\rangle$ in spatial modes b and c (up to a global phase). This is because the state $|\Phi^+\rangle$ has the following evolution:

$$
\begin{aligned}
|\Phi^+\rangle = &\ (|HH\rangle_{\mathrm{bc}} + |VV\rangle_{\mathrm{bc}})/\sqrt{2} \\
\xrightarrow{\text{BS 1}} &\ i\,(|HH\rangle_{\mathrm{b'b'}} + |VV\rangle_{\mathrm{b'b'}} + |HH\rangle_{\mathrm{c'c'}} + |VV\rangle_{\mathrm{c'c'}}))/2 \\
\xrightarrow{\text{EWPs \& EOMs}} &\ (|RR\rangle_{\mathrm{b'b'}} - |LL\rangle_{\mathrm{b'b'}} + |LL\rangle_{\mathrm{c'c'}} - |RR\rangle_{\mathrm{c'c'}})/2 \\
\xrightarrow{\text{BS 2}} &\ i(|HV\rangle_{\mathrm{b''b''}} - |HV\rangle_{\mathrm{c''c''}})/\sqrt{2};
\end{aligned}
$$

(b) If we detect a coincidence in different spatial modes b'' and c'' but with same polarization in the $|H\rangle/|V\rangle$ basis, photons 2 and 3 are projected onto $|\Phi^-\rangle$ in spatial modes b and c (up to a global phase). This is because the state $|\Phi^-\rangle$ has the following evolution:

$$
\begin{aligned}
|\Phi^-\rangle = &\ (|HH\rangle_{\mathrm{bc}} - |VV\rangle_{\mathrm{bc}})/\sqrt{2} \\
\xrightarrow{\text{BS 1}} &\ i\,(|HH\rangle_{\mathrm{b'b'}} - |VV\rangle_{\mathrm{b'b'}} + |HH\rangle_{\mathrm{c'c'}} - |VV\rangle_{\mathrm{c'c'}}))/2 \\
\xrightarrow{\text{EWPs \& EOMs}} &\ (|RR\rangle_{\mathrm{b'b'}} + |LL\rangle_{\mathrm{b'b'}} + |LL\rangle_{\mathrm{c'c'}} + |RR\rangle_{\mathrm{c'c'}})/2 \\
\xrightarrow{\text{BS 2}} &\ i(|HH\rangle_{\mathrm{b''c''}} - |VV\rangle_{\mathrm{b''c''}})/\sqrt{2}
\end{aligned}
$$

Note that although it is possible to choose, via the QRNG, whether to project the quantum state of two photons onto an entangled or a separable state, it is not possible to choose onto which particular state. The outcome of the BSM (SSM) is random and hence one cannot have prior knowledge of whether one obtains $|\Phi^-\rangle_{23}$ or $|\Phi^+\rangle_{23}$ ($|HH\rangle_{23}$ or $|VV\rangle_{23}$) for each individual run. The experimental results of the correlation functions of photons 1 and 4 in the $|H\rangle/|V\rangle$, $|R\rangle/|L\rangle$ and $|+\rangle/|-\rangle$ bases are 0.589 ± 0.078, –0.561 ± 0.078, 0.59 ± 0.072, respectively. A state fidelity $F\big(\hat{\rho}_{\exp}, |\Phi^+\rangle_{14}\big)_{\mathrm{BSM}}$ of 0.685 ± 0.033 is obtained and the entanglement witness value $W\big(\hat{\rho}_{\exp}, |\Phi^+\rangle_{14}\big)_{\mathrm{BSM}}$ is –0.185 ± 0.033. Therefore, entanglement between photons 1 and 4 is verified.

When Victor performs a BSM, photons 1 and 4 are only entangled if there exists the information necessary for Victor to specify into which subensembles the data are to be sorted. In our case the subensembles correspond to $|\Phi^-\rangle_{23}$ or $|\Phi^+\rangle_{23}$. Without the ability for this specification, he would have to assign a mixture of these two Bell states to his output state which is separable, and thus he could not correctly sort Alice's and Bob's data into subensembles. This is confirmed by evaluating the experimental data obtained in a BSM but without discriminating between $|\Phi^-\rangle_{23}$ and $|\Phi^+\rangle_{23}$. Then there exists a correlation only in the $|H\rangle/|V\rangle$ basis (0.55 ± 0.06) and no correlations in the $|+\rangle/|-\rangle$ (0.02 ± 0.05) and $|R\rangle/|L\rangle$ (0.01 ± 0.05) bases, similar to the situation when Victor performs a separable-state measurement.

**Quantum random number generator**

Refs. (6, 21) of the main text describe the details of the working principle of our quantum random number generator (QRNG). A weak light beam from a light emitting diode splits on a balanced optical beam splitter. Quantum theory predicts that each individual photon, which is generated from the source and travels through the beam splitter, has the same probability for both output arms of the beam splitter. Photon multipliers (PM) are used to detect the photons on each output, labeled by '0' and '1'. When PM '0' fires, the bit value is set to 0. It remains 0 until PM '1' detects one photon, which flips the bit value to 1. Thus, the bit sequence produced by the QRNG is truly random according to quantum theory.

**Four-fold count rate and error estimation**

Photons 1 and 4 are filtered with two interference filters of 3 nm FWHM bandwidth, and photons 2 and 3 are filtered with 1 nm IFs. We have detected the 2-fold count rate directly (without tunable BiSA) of about 20 kHz, and about 4.9 Hz 4-fold count rate with a pump power of 700 mW. There are two reasons for the relatively low detected 4-fold coincidence counts. The first one is to avoid the higher order emissions from spontaneous parametric down conversion (SPDC). Therefore, we could not pump the crystal with too high pump power[2]. In this experiment, we just used 700 mW, half of the maximal pump power. The second reason is the loss on various optical components and the duty cycle of the EOMs. We lost 79% of the photons on each individual input of the tunable BiSA (including the 104 m single mode fiber), which results in 4.4% photons left. The probabilistic nature of the Bell-state projection with linear optics decreases the success probability to 1/4. The random choices to either perform BSM or SSM determined by the QRNG gives an additional trivial reduction of 1/2 for each individual measurement. As explained above, the duty cycle of the EOMs are not unity, which reduces the count rate to 0.6. Including all of these losses, we are left with the fraction 0.0033 of the initial 4.9 Hz for each individual measurement. Therefore, the detected four-fold rate is about 0.016 Hz, as stated in the caption of Fig. 2 of the main text.

The limited state fidelities are mainly due to the higher order emissions from SPDC, although we have deliberately reduced the UV pump power. We have developed a numerical model to calculate the expected results. This model is based on the interaction Hamiltonian of SPDC. Given an interaction strength (or squeezing parameter) of the pump and nonlinear crystal, which can be measured experimentally, one can expand the Hamiltonian into a Taylor series. From this expansion, we can estimate the noise from the higher order emission. We utilized a quantum optics toolbox in Matlab[3], based on a matrix representation of quantum states with up to three photons per mode. With this model, it is straightforward to reproduce the count rates and visibilities of our system. In the model the detection efficiency is determined by the specifications of the single photon detectors and the coupling efficiency is measured and derived from the ratio of the 2-fold coincidence counts and single counts. From this calculation, the expected correlation function of photons 1 and 4 is about 0.674.

The other reason which limits the state fidelities is the group velocity mismatch of pump photons and the down converted photons in the type-II phase matching of BBO crystal[4]. A rigorous model for calculating that can be found in Refs. (5,6). The expected correlation function of photons 1 and 4 decreases to 0.964.

We are also limited by experimental imperfections. For instance, the performance of the tunable BiSA, which is limited by the visibility of the MZI (0.95) and the switching fidelity (0.99), is about 0.94. The polarization alignment in the fibers, which quantifies the transmission fidelity of the photon in the fibers, is about 0.99.

The overall expected correlation function of photons 1 and 4 is the product of all above listed values and equals to 0.605, which is in good agreement with our measured value.

### Experimental fulfillment of the delayed-choice condition

The experimental scheme and the time diagram of the relevant events is shown in Fig. 1. We assigned that the generation of photons 1 and 2 (event $\mathbf{G_I}$) happened at 0 ns, as the origin of the diagram. The generation of photons 3 and 4 (event $\mathbf{G_{II}}$) happened 1.6 ns later. At 35 ns, the measurements of Alice and Bob (events $\mathbf{M_A}$ and $\mathbf{M_B}$) occurred. The choice of Victor (event $\mathbf{C_V}$) was made by the QRNG in the time interval ranging from 49 ns to 348 ns and sent to the tunable BiSA. Due to the fibre delay of photons 2 and 3, at 520 ns Victor performed the bipartite state measurement (event $\mathbf{M_V}$) according to the bit value of his choice. Note that our definition of the choice event is very conservative. This is because in addition to the fixed amount of the electrical delay of the EOMs' driver (45 ns), QRNG (75 ns) and connecting cables (20 ns), we also included 3 times the QRNG autocorrelation time (3·10.7 ns ≈ 32 ns) and the on-time of the EOMs (299 ns). This on-time gave the time of event $\mathbf{C_V}$ a lower bound of 49 ns and an upper bound of 348 ns. As shown in Fig. 1, it is clear to see for each successful run (a 4-fold coincidence count) that not only event $\mathbf{M_V}$ happened 485 ns later than events $\mathbf{M_A}$ and $\mathbf{M_B}$, but also event $\mathbf{C_V}$ happened 14 ns to 313 ns later than events $\mathbf{M_A}$ and $\mathbf{M_B}$ even in this conservative consideration. Therefore, with this configuration we unambiguously fulfilled the delayed-choice condition. Note that the main uncertainty of our experiment in time for measurements is the detector jitter, which is about 800 ps.

## References

1. X.-S. Ma *et al.,* A high-speed tunable beam splitter for feed-forward photonic quantum information processing. *Opt. Express*, **19**, 22723-22730 (2011).
2. A. Lamas-Linares, J. C. Howell, D. Bouwmeester, Stimulated emission of polarization-entangled photons. *Nature*, **412**, 887-890 (2001).
3. S.-M. Tan, A computational toolbox for quantum and atomic optics. *J. Opt. B: Quantum Semiclass. Opt.* **1**, 424 (1999).
4. P. Mosley *et al.*, Heralded Generation of Ultrafast Single Photons in Pure Quantum States. *Phys. Rev. Lett.* **100**, 133601 (2008).
5. T. Jennewein, R. Ursin, M. Aspelmeyer, A. Zeilinger, Performing high-quality multi-photon experiments with parametric down-conversion. *J. Phys. B: At. Mol. Opt. Phys.* **42**, 114008 (2009).
6. R. Kaltenbaek, Entanglement swapping and quantum interference with independent sources. Ph.D. thesis, University of Vienna (2008).

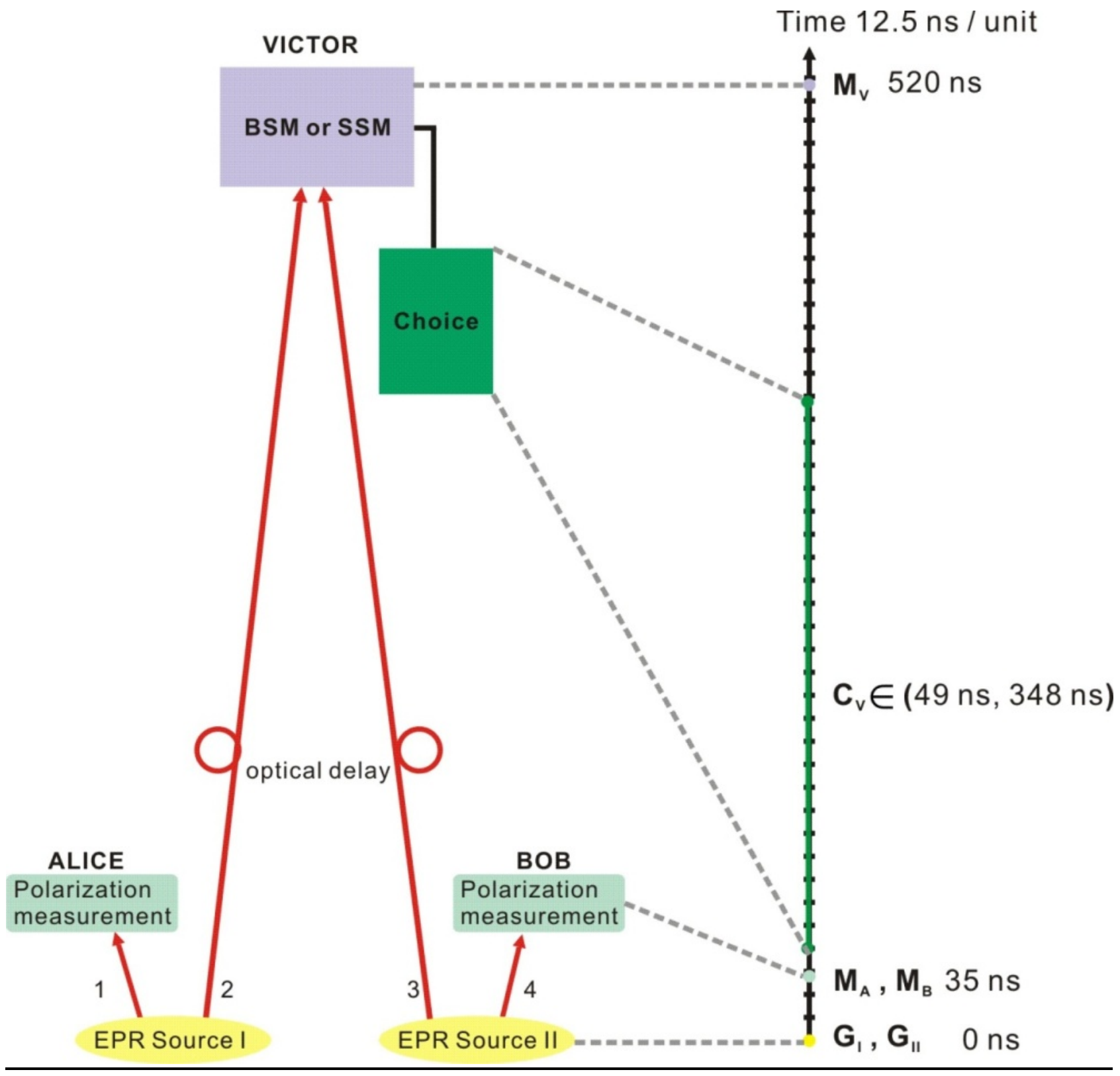

Figure 1: Time diagram of our delayed-choice entanglment swapping experiment. Two polarization entangled photon pairs (1&2 and 3&4) are generated from Einstein-Podolsky-Rosen (EPR) sources I and II (events $\mathbf{G_I}$ and $\mathbf{G_{II}}$) at 0 ns. The polarizations of photons 1 and 4 are measured by Alice (event $\mathbf{M_A}$) and Bob (event $\mathbf{M_B}$) 35 ns later. The other two photons (photons 2 and 3) are delayed and then sent to Victor who can choose (event $\mathbf{C_V}$) to swap the entanglement or not by performing a Bell-state measurement (BSM) or a separable-state measurement (SSM) on photons 2 and 3 (event $\mathbf{M_V}$). Victor's choice and measurement are made after Alice and Bob's polarization measurements. Remarkably, whether the earlier registered results of photons 1 and 4 indicate the existence of entanglement between photons 1 and 4 depends on the later choice of Victor.